\documentclass[aps,prd,preprint,showpacs,superscriptaddress]{revtex4}
\usepackage{amsmath}
\usepackage{latexsym}
\usepackage{amsfonts}
\usepackage{graphicx}
\usepackage{mathrsfs}
\usepackage{epstopdf}
\usepackage{color}
\usepackage[colorlinks,linkcolor=magenta,anchorcolor=blue,citecolor=blue]{hyperref}

\def\be{\begin{equation}}
\def\ee{\end{equation}}
\def\bea{\begin{eqnarray}}
\def\eea{\end{eqnarray}}

\begin{document}
\title{Confronting the Potential-driven $D$BI-$i$nspired $n$onminimal $ki$netic $c$oupling (Dinkic) inflation to the observational data}

\author{Jun Chen}
\email{junchen@mails.ccnu.edu.cn}
\affiliation{Key Laboratory of Quark and Lepton Physics (MOE), Central China Normal University, Wuhan, Hubei 430079, China}
\affiliation{Institute of Particle Physics, Central China Normal University, Wuhan, Hubei 430079, China}
\affiliation{Institute of Astrophysics, Central China Normal University, Wuhan, Hubei 430079, China}

\author{Hou Wenjie}
\email{houwenjie@mails.ccnu.edu.cn}
\affiliation{Key Laboratory of Quark and Lepton Physics (MOE), Central China Normal University, Wuhan, Hubei 430079, China}
\affiliation{Institute of Particle Physics, Central China Normal University, Wuhan, Hubei 430079, China}
\affiliation{Institute of Astrophysics, Central China Normal University, Wuhan, Hubei 430079, China}

\author{Defu Hou}
\email{houdf@mail.ccnu.edu.cn}
\affiliation{Key Laboratory of Quark and Lepton Physics (MOE), Central China Normal University, Wuhan, Hubei 430079, China}
\affiliation{Institute of Particle Physics, Central China Normal University, Wuhan, Hubei 430079, China}

\author{Taotao Qiu}
\email{qiutt@mail.ccnu.edu.cn}
\affiliation{Key Laboratory of Quark and Lepton Physics (MOE), Central China Normal University, Wuhan, Hubei 430079, China}
\affiliation{Institute of Astrophysics, Central China Normal University, Wuhan, Hubei 430079, China}

\date{\today}

\begin{abstract}
	
In the previous work \cite{Qiu:2015aha}, a new kind of inflation model was proposed, which has the interesting property that its perturbation equation of motion gets a correction of $k^4$, due to the non-linearity of the kinetic term. Nonetheless, the scale-invariance of the power spectrum remains valid, both in large-$k$ and small-$k$ limits. In this paper, we investigate in detail the spectral index, the index running and the tensor/scalar ratio in this model, especially on the potential-driven case, and compare the results to the current PLANCK/BICEP observational data. We also discuss the tensor spectrum in this case, which is expected to be tested by the future observations on primordial gravitational waves.

\end{abstract}

\maketitle
\section{Introduction}
Along with the development of the theory of inflationary cosmology, more and more interesting inflation models are coming out, with various theoretical motivations and/or observational advantages. Among them are Galileon/Horndeski models \cite{Horndeski:1974wa, Nicolis:2008in, Deffayet:2009mn} which were proposed for last several decades. In this theory, higher derivative terms and/or nonminimal couplings are included, however the interesting property is that there is no (bad) redundant dynamical degrees of freedom, which is due to the delicate design of the action. 

In order to extend this theory to explore more interesting properties, people also become keen on constructing Beyond Horndeski models \cite{Appleby:2012rx, Gleyzes:2014dya, Ohashi:2015fma, Langlois:2015cwa}. In the previous work \cite{Qiu:2015aha}, one of the current authors proposed a new kind of inflation model which contains a nonminimal kinetic coupling term, and moreover, by using a DBI-type form of the action, the correction term is nonlinearly included in the action. We thus dub this model as "DBI-inspired nonminimal kinetic coupling" (Dinkic) inflation model. This model shares nice ghost-free property of Galileon/Horndeski models, however, there will be a correction term which is proportional to $k^4$ in the perturbed equation of motion, due to the nonlinear form. Nevertheless, the scale invariance of the power spectrum can still be garanteed, even in the large-$k$ limit when this term is dominant, but due to the coupling, there is an deficit of the power spectrum in the large-$k$ limit which makes the whole spectrum red-tilted.

On the other hand, since the observations also develops with more and more precision and detectability, we expect to get more and more information from the observational data. Apart from the most frequently used parameters such as the amplitude of primordial power spectrum $P_\zeta$, the spectral index $n_s$ and the tensor/scalar ratio $r$, we have also been able to detect more details of the spectrum such as the running of the index ($\alpha_s\equiv dn_s/d\ln k$) and even the running of the running ($\beta_s\equiv d^2n_s/d\ln k^2$). For example, the PLANCK 2013 data has put stringent constraint on the $\alpha_s$: $\alpha_s=-0.013\pm0.009$ (68\% C.L.) \cite{Planck:2013jfk}, while in the PLANCK 2015 data, the constraints are improved to be $\alpha_s=-0.0033\pm0.0074$ (68\%, C.L., $\beta_s=0$), or $\alpha_s=0.009\pm0.010$, $\beta_s=0.025\pm0.013$ (68\%, C.L., $\beta_s\neq0$) \cite{Ade:2015lrj}. 

Besides the scalar parts, it becomes more and more important to pay attention to the tensor modes of primordial perturbations, because they can generate primordial gravitational waves \cite{Zaldarriaga:1996xe}. The recent discovery of binary gravitational waves (GW) \cite{Abbott:2016blz} won the Nobel Prize and becomes big deal all over the world, however the next goal of detecting gravitational waves with lower frequency is a more challenging. There will be lots of world-class projects for low-frequency GW, such as eLISA Satellite \cite{AmaroSeoane:2012km}, TianQin Satellite \cite{Luo:2015ght}, and AliCPT Telescope \cite{Li:2017drr}. The AliCPT telescope (first stage), located in the Ali region of Tibet, China with an altitude of 5,250 meters, is expected to improve the constraint on the tensor/scalar ratio to about one order of magnitude stronger within a few years. One may also be able to extract more information about tensor perturbations, such as the tensor spectral index $n_T$ and so on \cite{Li:2017drr}.  

In this paper, we will perform more detailed investigation of the Dinkic inflation model, paying special attention to the case in which inflation is driven by the potential of the model. There have been many famous potential-driven inflation models such as chaotic inflation \cite{Linde:1983gd}, natural inflation \cite{Freese:1990rb}, new inflation \cite{Albrecht:1982wi, Albrecht:1984qt}, axion monodromy inflation \cite{Silverstein:2008sg} and so on, some of which have fundamental origin or close related to particle physics  (see comprehensive review \cite{Martin:2013tda}). 
In the following, by considering various inflation potentials, we will calculate the power spectrum, spectral index and its running, and confront with the observational data. Moreover, we also consider its information about tensor perturbations such as tensor spectral index and tensor/scalar ratio, and compare them to observational constraints from not only current data, but also future detectors of primordial gravitational waves. Note that the similar consideration for canonical inflation with nonminimal kinetic coupling has been performed in Ref. \cite{Yang:2015pga}.

Our paper is organized as follows: in Sec. \ref{Dinkic} we briefly review the Dinkic inflation model, and present general analytical forms of scalar and tensor spectra, spectral index and its running, as well as the tensor/scalar ratio. In Sec. \ref{nsasr}, for various parameter choices of large field and small field models, we get solutions of quantities for scalar perturbations given in Sec. \ref{Dinkic}, and their observational constraints. In Sec. \ref{ntr} we discuss the tensor perturbations of these models. Sec. \ref{conclusion} attributes to concluding remarks. 

\section{The Dinkic Inflation model}
\label{Dinkic}
To start with, we write down the action of the Dinkic inflation model as \cite{Qiu:2015aha}:
\be\label{action}
S=\int d^{4}x\sqrt{-g}\left[\frac{R}{2\kappa^{2}}-\frac{1}{f(\phi)}(\sqrt{\mathcal{D}}-1)-V(\phi)\right]~,
\ee
where $\mathcal{D}\equiv1-2\alpha f(\phi)X+2\beta f(\phi)\widetilde{X}$, with $X\equiv-\frac{1}{2}g^{\mu\nu}\partial_{\mu}\phi\partial_{\nu}\phi$ and $\widetilde{X}\equiv-\frac{1}{2M^{2}}G^{\mu\nu}\partial_{\mu}\phi\partial_{\nu}\phi$. $M$ is the scale of nonminimal kinetic coupling, while $M_{pl}=\kappa^{-1}$ is the Planck scale. Note that since the kinetic coupling term $2\beta f(\phi)\widetilde{X}$ resides in the square root, the action becomes non-linear and cannot be included in the Galileon or Horndeski theories. Under the flat FRW metric ($g_{\mu\nu}=diag\{-1,a^2(t),a^2(t),a^2(t)\}$) where $X=\dot\phi^2/2$, $\widetilde{X}=-3(H/M)^2\dot\phi^2/2$, by varying the action with respect to the field $\phi$, we can get the equation of motion for $\phi$:
\be\label{eom}
0=\frac{f_{\phi}(\sqrt{\mathcal{D}}-1)^{2}}{2f^{2}\sqrt{\mathcal{D}}}+\frac{3\beta H^{2}-\alpha}{\sqrt{\mathcal{D}}}\ddot{\phi}+\frac{2\beta\dot{H}+3\beta H^{2}-\alpha}{\sqrt{\mathcal{D}}}3H\dot{\phi}-\frac{3\beta H^{2}-\alpha}{2\mathcal{D}^{3/2}}\dot{\mathcal{D}}\dot{\phi}-V_{\phi}~, 
\ee
and the energy density $\rho$ and pressure $p$ are:
\bea
\label{rho}
\rho&=&\frac{(\sqrt{\cal D}-1)}{f(\phi)}+V(\phi)+\frac{\alpha\dot{\phi}^{2}}{\sqrt{\mathcal{D}}}+\frac{6\beta H^{2}\dot{\phi}^{2}}{M^2\sqrt{\mathcal{D}}}~,\\
\label{p}
p&=&-\frac{(\sqrt{\cal D}-1)}{f(\phi)}-V(\phi)-\frac{3\beta H^{2}\dot{\phi}^{2}}{M^2\sqrt{\mathcal{D}}}-\left(\frac{\beta H\dot{\phi}^{2}}{M^2\sqrt{\mathcal{D}}}\right)~,
\eea
which satisfy the Friedmann Equations $3H^2=\kappa^2\rho$, $\dot H=-\kappa^2(\rho+p)/2$. From the above equations one can see that, in the absence of the $2\beta\tilde X$ term and in the slow-roll limit ($\alpha f \dot\phi^2\ll 1$, $\sqrt{\cal D}\sim 1$), everything returns to the slow-roll canonical inflation case. In the presence of the $2\beta\tilde X$ term and in the slow-roll limit ($|\alpha f \dot\phi^2-3 \beta f (H/M)^2\dot\phi^2|\ll 1$, $\sqrt{\cal D}\sim 1$), everything will be the same as the slow-roll canonical inflation with kinetic term nonminimally couple to the Einstein tensor in background level. However, as will be shown below, due to the non-linearity of the nonminimal coupling term, the perturbation equation of motion will have a $k^4$-correction term which is different from its canonical correspondence. This will also cause some difference in the observables in the small scale (large $k$) region.  

To analyse the perturbations generated by the model, we make use of the Arnowitt-Deser-Misner (ADM) formalism \cite{Arnowitt:1962hi}. 
The perturbed action up to the second order becomes \cite{Qiu:2015aha}:
\be\label{2ndaction}
\delta^2S\approx\frac{1}{2\kappa^{2}}\int d^{4}xa^{3}\Big[6\frac{x_\beta}{\mathcal{D}}\dot{\zeta}^{2}-\frac{2\epsilon}{a^{2}}(\partial\zeta)^{2}+\frac{16x_\beta^{4}y}{a^{4}H^{2}}(\partial^{2}\zeta)^{2}\Big]~,
\ee
where $x_\beta\equiv\kappa^{2}\beta\dot{\phi}^{2}/(2M^2\sqrt{\mathcal{D}})$, $y\equiv f(\phi)M_{p}^2H^{2}/\sqrt{\mathcal{D}}$, and $\epsilon\equiv-\dot H/H^2$. Moreover, to get rid of the ghost problem, $x_\beta$ is required to be larger than zero.

From the action, one can easily get the perturbed equation of motion:
\be\label{perturbeomscalar2}
u^{\prime\prime}+c_s^2k^2\left[1+24\frac{x_\beta^5|y|}{\epsilon^2{\cal D}^2}\left(\frac{c_sk}{aH}\right)^2\right]u-\frac{z^{\prime\prime}}{z}u=0~,
\ee
where $u\equiv z\zeta$, $z\equiv a\sqrt{3x_\beta/{\cal D}}$, $c_s^2=\epsilon{\cal D}/3x_\beta$, and prime denotes derivative with respect to conformal time $\tau\equiv\int a^{-1}(t) dt$. The solution is:
\be\label{perturbeom}
u=\frac{\sqrt{\pi|\tau|}}{2}[H^{(1)}_{\nu}(\omega\tau)+H^{(1)}_{-\nu}(\omega\tau)]~,
\ee
with $H^{(1)}$ being the type I Hankel function, and we approximately have 
\be
\nu\simeq 3\int\omega d\eta/(2\omega\tau)~,  \qquad \omega^2=\frac{\epsilon\mathcal{D}}{3x_\beta}[1+24\frac{x^5_\beta |y|}{\epsilon^2\mathcal{D}^2}\big(\frac{c_s k}{aH}\big)^2]~.
\ee

In the large scale limit where $k^{-1}\gg k_c^{-1}$ by a critical $k$-value $k_c\equiv aH\sqrt{\epsilon \mathcal{D}/(8x_\beta^4 y)}$, we have $\nu\simeq 3/4$, while in the small scale limit where $k^{-1}\ll k_c^{-1}$, one has $\nu\simeq3/2$.

The superhorizon solution of Eq. (\ref{perturbeom}) is
\be
\zeta \simeq \frac{1}{3} \int \frac{{\cal D}dt}{a^3(t)x_\beta}~~~(large~scale)~~or~~H\sqrt{\frac{{\cal D}}{6x_\beta\omega^3}}~~~(small~scale)~,
\ee
so that in large scale limit, the power spectrum, the spectral index and the running of spectral index are:
\bea
\label{ps}
P^{(l)}_S&\equiv&\frac{k^3}{2\pi^2}|\zeta|^2\simeq\frac{H^2}{8\pi^2}\sqrt{\frac{3x_\beta}{\epsilon^3{\cal D}}}~,\\
\label{nss}
n^{(l)}_S&\equiv&1+\frac{d\ln P^{(l)}_S}{d\ln k}\simeq1-2\epsilon-\frac{3}{2}\eta+\iota-\frac{3}{4}s~,\\
\label{ass}
\alpha^{(l)}_S&\equiv&\frac{dn^{(l)}_{S}}{d\ln k}\simeq-2\eta\epsilon-\frac{3}{2}h \eta+j \iota-\frac{3}{4}\varsigma s~,
\eea
respectively, where $\eta\equiv\dot\epsilon/(H\epsilon)$, $s\equiv\dot{\mathcal{D}}/(H\mathcal{D})$, $\iota\equiv\ddot{\phi}/(H\dot{\phi})$, $h\equiv\dot\eta/(H\eta)$, $j\equiv\dot\iota/(H\iota)$, $\varsigma\equiv\dot s/(Hs)$. While in the small scale limit where the $k^4$-term is taken into account, the power spectrum, the spectral index and the running of spectral index become:
\bea
\label{pl}
P^{(s)}_S&\simeq&\frac{H^2}{8\pi^2}\sqrt{\frac{3x_\beta}{\epsilon^3{\cal D}}}\left[1-{\cal C}\left(\frac{c_sk}{aH}\right)^2\right]~,\\
\label{nsl}
n^{(s)}_S&\equiv&1+\frac{d\ln P^{(s)}_S}{d\ln k}\simeq1-2\epsilon-\frac{3}{2}\eta+\iota-\frac{3}{4}s-\frac{{\cal C}}{1-{\cal C}}(5\epsilon_x+\epsilon_y-2\eta-2s)~,\\
\label{asl}
\alpha^{(s)}_S&\equiv&\frac{dn^{(s)}_{S}}{d\ln k}\simeq-2\eta\epsilon-\frac{3}{2}h \eta+j \iota-\frac{3}{4}\varsigma s-\frac{{\cal C}}{(1-{\cal C})^{2}}(5\epsilon_{x}+\epsilon_{y}-2\eta-2s)^2 \nonumber\\
&&-\frac{{\cal C}}{1-{\cal C}}(-10j \iota+8\eta\epsilon-2h\eta+\iota\epsilon_f+\epsilon_f\epsilon+\frac{\epsilon_f^2}{4})~,
\eea
where $\epsilon_x\equiv\dot x_\beta/(Hx_\beta)$, $\epsilon_y\equiv\dot y/(Hy)$, $\epsilon_f\equiv\dot{f}/Hf$, and ${\cal C}\equiv36x_{\beta}^{5}|y|/\epsilon^{2}\mathcal{D}^{2}$. One can see that in both limit the power spectrum can be nearly scale-invariant, but in the small scale limit, there is a correction to the amplitude of the power spectrum, which is due to the $k^4$ term correction in the expression of $\omega^2$. 

Moreover, one can also calculate the tensor perturbation by considering the tensor mode of gravitational fluctuations. The second-order action for tensor perturbation can be written as \cite{Qiu:2015aha}:
\be 
\delta^2S^T=\frac{1}{8\kappa^2}\int d^4xa^3\left[{\cal F}_T\dot\gamma_{ij}^2-{\cal G}_T\frac{(\nabla\gamma_{ij})^2}{a^2}\right]~,
\ee
where ${\cal F}_T\equiv1-x_\beta$, ${\cal G}_T\equiv1+x_\beta$. The equation of motion for tensor perturbation $\gamma_{ij}$ is
\be
\gamma_{ij}^{\prime\prime}-c_T^2\nabla^2\gamma_{ij}+\frac{(a^2{\cal F}_T)^\prime}{a^2{\cal F}_T}\gamma_{ij}^\prime=0~,
\ee
where $c_T^2\equiv{\cal G}_T/{\cal F}_T$. The above equation has the solution:
\be
\gamma_{ij}=C_{1}+C_{2}\int\frac{dt}{a^3(t){\cal F}_T}~,
\ee
where $C_1$ and $C_2$ are integration constants. The power spectrum and the spectral index for tensor perturbation is 
\bea
P_{T}&\equiv&\frac{k^3}{2\pi^2}|\gamma_{ij}|^2=\frac{2H^{2}}{{\cal G}_T c_{T}\pi^{2}}~,\\
\label{nt}
n_{T}&\equiv&\frac{d\ln P_T}{d\ln k}=\frac{x_\beta}{x_\beta-1}(2\iota-s)-2\epsilon-s_{T}~,
\eea
where $s_{T}\equiv\dot{c}_{T}/(Hc_{T})$, and with Eqs. (\ref{ps}) and (\ref{pl}), one gets the tensor/scalar ratio $r$ as:
\bea
\label{rs}
r^{(l)}&\equiv&\frac{P_T}{P^{(l)}_S}\simeq 16\epsilon\sqrt{\frac{\epsilon{\cal D}}{3x_\beta}}~,\\
\label{rl}
r^{(s)}&\equiv&\frac{P_T}{P^{(s)}_S}\simeq 16\epsilon\sqrt{\frac{\epsilon{\cal D}}{3x_\beta}}\left[1+{\cal C}\left(\frac{c_sk}{aH}\right)^2 \right]~,
\eea
for large and small scale limits, respectively.

\section{Spectrum, index and its running}
\label{nsasr}
\subsection{Slow-roll Analysis in Potential-driven Case}
In this section, we restrict ourselves to a specific case, where the inflation is driven by the potential. While in this case slow-roll approximation can be applied, we consider the following slow-roll conditions:
\be\label{srcon}
\frac{1}{2}\alpha\dot{\phi}^{2}+\frac{9\beta H^{2}\dot{\phi}^{2}}{2M^{2}}\ll V(\phi)~,~|\ddot{\phi}|\ll |3H\dot{\phi}|~,~\frac{|2\beta\dot{H}|}{|\alpha M^{2}+3\beta H^{2}|}\ll 1~,
\ee
under which the equation of motion (\ref{eom}) and Friedmann equation are reduced to:
\be\label{eomsr}
3(\alpha+\frac{3\beta H^{2}}{M^{2}})H\dot{\phi}+V_{\phi}\simeq 0~,~\frac{3H^{2}}{\kappa^{2}}\simeq V(\phi)~,
\ee
and the e-folding number $N$ is defined as:
\be\label{efold}
N\equiv\int_{t_\ast}^{t_e} Hdt=\int_{\phi_\ast}^{\phi_e}\frac{H}{\dot\phi}d\phi\simeq-\int_{\phi_\ast}^{\phi_e}\frac{\kappa^2V(\alpha M^2+\beta\kappa^2V)}{M^2V_{\phi}}d\phi~.
\ee

It is useful to define the potential-based slow-roll parameters as:
\be\label{srpara}
\epsilon_{V}=\frac{M^2}{6\kappa^{2}}\Big(\frac{V_{\phi}}{V}\Big)^{2}\frac{(\alpha M^2+3\beta\kappa^2V)}{(\alpha M^2+\beta\kappa^2V)^{2}}~,~\eta_{V}=\frac{M^2V_{\phi\phi}}{(\alpha M^2+\beta\kappa^2V)\kappa^{2}V}~,~\xi_V^2=\frac{M^4V_{\phi\phi\phi}V_{\phi}}{(\alpha M^2+\beta\kappa^2V)^{2}\kappa^{4}V^{2}}~.
\ee
In high friction limit where $3\beta H^2\simeq\beta\kappa^2V\gg\alpha M^2$, one can reduce the above formulae to get: $N\simeq-\beta\kappa^4/M^2\int_{\phi_\ast}^{\phi_e}(V^2/V_{\phi})d\phi$, $\epsilon_V\simeq M^2V_{\phi}^2/(2\kappa^{4}\beta V^3)$, $\eta_V\simeq M^2V_{\phi\phi}/(\beta\kappa^4V^2)$, $\xi_V^2\simeq M^4V_{\phi\phi\phi}V_{\phi}/(\beta^2\kappa^8V^4)$, which is different from the canonical single scalar field models. By using the Eq. (\ref{eomsr}), the geometry-based slow-roll parameters ($\epsilon$, $\eta$, $\iota$, $s$, $h$, $j$, $\varsigma$) can be re-expressed as: 
\bea\label{relation}
&&\epsilon\simeq\epsilon_{V}~,~\eta\simeq 2\iota\simeq 2(3\epsilon_{V}-\eta_{V})~,~s\simeq 4\iota-2\epsilon\simeq 10\epsilon_{V}-4\eta_{V}~,\nonumber\\
&&h\simeq j\simeq\frac{18\epsilon_V^2+\xi_V^2-10\epsilon_V\eta_V}{3\epsilon_V-\eta_V}~,~\varsigma\simeq\frac{30\epsilon_V^2-18\epsilon_V\eta_V+2\xi_V^2}{5\epsilon_V-2\eta_V}~.
\eea
Moreover, $x_\beta$ can also be expressed as $x_{\beta}\simeq\epsilon_{V}/3$. Therefore, in the large scale limit, according to Eqs. (\ref{nss}), (\ref{ass}) and (\ref{rs}), one can express $n_s$, $\alpha_s$ and $r$ as:
\bea
\label{nslsr}
n_{s}^{(l)}&\simeq&1-\frac{31}{2}\epsilon_{V}+5\eta_{V}~,\\
\label{aslsr}
\alpha^{(l)}_s&\simeq& -93\epsilon_{V}^2+51\epsilon_{V}\eta_{V}-5\xi^2_V~,\\
\label{rlsr}
r^{(l)}&\simeq&16\epsilon_V~.
\eea
In the small scale limit, some more parameters need to be taken into account. From the expressions of $\epsilon_x$ and $\epsilon_y$, we have:
\be
\epsilon_{x}\simeq 2\iota-\frac{1}{2}s~,~~~\epsilon_{y}\simeq\epsilon_f-2\epsilon-\frac{1}{2}s~.
\ee
Therefore, using (\ref{nsl}), (\ref{asl}) and (\ref{rl}),  $n_s$, $\alpha_s$ and $r$ turn out to be:
\bea
\label{nsssr}
n_{s}^{(s)}&\simeq&1-\frac{31}{2}\epsilon_{V}+5\eta_{V}+\frac{{\cal C}}{1-{\cal C}}(34\epsilon_{V}-14\eta_{V}-\epsilon_{f})~,\\ 
\label{asssr}
\alpha^{(s)}_s&\simeq&-93\epsilon_{V}^{2}+51\epsilon_{V}\eta_{V}-5\xi_{V}^{2}+\frac{\cal C}{1-{\cal C}}(204\epsilon_{V}^{2}-124\epsilon_{V}\eta_{V}+14\xi_{V}^{2}+\eta_{V}\epsilon_{f}-4\epsilon_{V}\epsilon_{f}-\frac{\epsilon_{f}^{2}}{4}) \nonumber\\
&&-\frac{\cal C}{(1-{\cal C})^{2}}(34\epsilon_{V}+14\eta_{V}-\epsilon_{f})^{2}~, \\
\label{rssr}
r^{(s)}&\simeq&16\epsilon_{V}(1+{\cal C})~.
\eea

One can see that, in the slow-roll approximation, these parameters of potential-driven Dinkic inflation model can be expressed with less slow-roll parameters and moreover, more analyzable. In the following, we will perform the calculation of the spectral index and its running, by taking as examples two typical cases of potential, namely the ``large field potential" and the ``small field potential", to see if the potential-driven Dinkic inflation model can be consistent with the observational data. 

\subsection{Inflation with Large Field Potential}

In the large field inflation models, the inflaton field goes from a large value towards a small value. This kind of models can usually give rise to an ``attractor" behavior of inflation without fine-tuning of the initial conditions \cite{Linde:1983gd, Brandenberger:2016uzh}, and also a large tensor/scalar ratio \cite{Linde:1983gd}. A commonly-used potential of large-field model is 
\be\label{largeV}
V(\phi)=\lambda M_{pl}^4(\phi/M_{pl})^n~,~~~n>0~. 
\ee
Note that for various index $n$, the potential (\ref{largeV}) can be reduced to various interesting examples. For $n=2$, (\ref{largeV}) is reduced to the mass-squared potential ($V=m^2\phi^2/2$) where $m=\sqrt{2\lambda}/M_p$ is the mass of the inflaton field, while for $n=4$, (\ref{largeV}) is reduced to chaotic inflation \cite{Linde:1983gd}. Moreover, $n$ can even be a rational number, for example,  for $n=2/3$ (\ref{largeV}) turns to the potential of axion monodromy inflation \cite{Silverstein:2008sg} where the inflaton field is considered to be reduced from a $D4$ brane action wrapped on compact manifold.

From Eq. (\ref{srpara}) and with the form of potential (\ref{largeV}), one can get the potential-based slow-roll parameter as:
\be\label{srparal}
\epsilon_{V}=\frac{n^{2}M^{2}M_{pl}^{n}}{2\beta\lambda\phi^{n+2}}~,~\eta_{V}=\frac{n(n-1)M^{2}M_{pl}^{n}}{\beta\lambda\phi^{n+2}}~,~\xi_V^2=\frac{n^{2}(n-1)(n-2)M^{4}M_{pl}^{2n}}{\beta^{2}\lambda^{2}\phi^{2n+4}}~
\ee
which are in pretty consistency with Ref. \cite{Yang:2015pga}. This is because the slow-roll parameters are constructed of background quantities, and as is mentioned in Sec. \ref{Dinkic}, the current model can be reduced to the model in Ref. \cite{Yang:2015pga} in background level. Even though, as will be shown below, due to their difference in perturbation level, the perturbation quantities such as $n_s$, $\alpha_s$ and $r$ are quite different.

Moreover, from Eq. (\ref{efold}), one can get the e-folding number of inflation from the horizon-crossing to the end:
\be\label{Nl}
N=\int_{\phi_{\ast}}^{\phi_{e}}\left(-\frac{\beta\lambda}{n}\frac{\phi^{n+1}}{M^2M_{pl}^{n}}\right)d\phi=\frac{\lambda\beta(\phi_{*}^{n+2}-\phi_{e}^{n+2})}{M^2 M_{pl}^{n}n(n+2)}~,
\ee
where $\phi_e$ and $\phi_\ast$ are the values of the field when inflation ends and when the perturbation of inflation observed today begins to cross the horizon, respectively.

For $0<n<2$, $\epsilon_V$ will reach unity earlier than $\eta_V$ (also shown in \cite{Yang:2015pga}), so we set the time when $\epsilon_V=1$ to be the ending time of inflation. Thus the final value of $\phi$ at the ending of inflation is obtained as:   
\be
\label{phiel1}
\phi_e=\left[\frac{n^2M^2M_{pl}^n}{2\beta\lambda}\right]^{1/(n+2)}~,
\ee
and Eqs. (\ref{phiel1}) and (\ref{Nl}) gives the solution:
\be
\label{phifl1}
\phi_\ast=\left[\frac{M_{pl}^{n}M^{2}n(n+2)}{\beta\lambda}\left(N+\frac{n}{2(n+2)}\right)\right]^{1/(n+2)}~.
\ee

Therefore, the slow-roll parameters at the crossing point turns out to be:
\bea
\epsilon_{V\ast}&=&\frac{n}{2N(n+2)+n}~,\\
\eta_{V\ast}&=&\frac{2(n-1)}{2N(n+2)+n}~,\\
\xi_{V\ast}^2&=&\frac{4(n-1)(n-2)}{[2N(n+2)+n]^{2}}~,
\eea
where here and after, we use subscript `$\ast$' to describe value at the crossing time. Moreover, one has $\epsilon_{f\ast}=8/[2N(n+2)+n]$. From Eqs. (\ref{nslsr})-(\ref{rlsr}) and (\ref{nsssr})-(\ref{rssr}), one gets the spectral index $n_s$, its running $\alpha_s$, and the tensor/scalar ratio $r$ as:
\bea
\label{nsllarge1}
n_{s\ast}^{(l)}&=&1-\frac{11n+20}{2[2N(n+2)+n]}~,\\
\label{asllarge1}
\alpha_{s\ast}^{(l)}&=&-\frac{(11n+20)(n+2)}{[2N(n+2)+n]^{2}}~,\\
\label{rllarge1}
r_\ast^{(l)}&=&\frac{16n}{2N(n+2)+n}~
\eea
for large scale limit, and
\bea
\label{nslsmall1}
n_{s\ast}^{(s)}&=&1-\frac{11n+20}{2[2N(n+2)+n]}+\frac{\cal C}{1-{\cal C}}\frac{6n+20}{2N(n+2)+n}~,\\
\label{aslsmall1}
\alpha_{s\ast}^{(s)}&=&-\frac{(11n+20)(n+2)}{[2N(n+2)+n]^2}+\frac{\cal C}{1-{\cal C}}\frac{4(3n+10)(n+2)}{[2N(n+2)+n]^2}\nonumber\\
&&-\frac{\cal C}{(1-{\cal C})^2}\frac{4(3n+10)^2}{[2N(n+2)+n]^2}~,\\
\label{rlsmall1}
r_\ast^{(s)}&=&\frac{16n(1+{\cal C})}{2N(n+2)+n}~
\eea
for small scale limit.

Oppositely, for $n\geq2$, the inflation ends when $\eta_V=1$ which reaches unity earlier than $\epsilon_V$. Therefore the value of $\phi$ at the ending of inflation could be determined as:   
\be
\label{phiel2}
\phi_{e}=\left[\frac{n(n-1)M_{pl}^{n}M^{2}}{\lambda\beta}\right]^{1/(n+2)}~.
\ee
Eqs. (\ref{phiel2}) and (\ref{Nl}) gives the solution:
\be
\label{phifl2}
\phi_\ast=\left[\frac{M_{pl}^{n}M^{2}n(n+2)}{\beta\lambda}\left(N+\frac{n-1}{n+2}\right)\right]^{1/(n+2)}~,
\ee
then the slow-roll parameters at the crossing point turn out to be:
\bea
\epsilon_{V\ast}&=&\frac{n}{2[N(n+2)+(n-1)]}~,\\
\eta_{V\ast}&=&\frac{(n-1)}{N(n+2)+(n-1)}~,\\
\xi_{V\ast}^2&=&\frac{(n-1)(n-2)}{[N(n+2)+(n-1)]^{2}}~,
\eea
and $\epsilon_{f\ast}=4/[N(n+2)+(n-1)]$. From Eqs. (\ref{nslsr})-(\ref{rlsr}), one gets the spectral index $n_s$, its running $\alpha_s$, and the tensor/scalar ratio $r$ as:
\bea
\label{nsllarge2}
n_{s\ast}^{(l)}&=&1-\frac{11n+20}{4[N(n+2)+(n-1)]}~,\\
\label{asllarge2}
\alpha_{s\ast}^{(l)}&=&-\frac{(11n+20)(n+2)}{4[N(n+2)+(n-1)]^{2}}~,\\
\label{rllarge2}
r_\ast^{(l)}&=&\frac{8n}{N(n+2)+(n-1)}~,
\eea
in large scale limit. For various choices of $n=2/3, 2, 4$ and $N_\ast=50, 60$, we list the specific values of  $n_s$, $\alpha_s$ and $r$ in Table \ref{largelT} respectively. Moreover, from (\ref{nsssr})-(\ref{rssr}), we get: 
\bea
\label{nslsmall2}
n_{s\ast}^{(s)}&=&1-\frac{11n+20}{4[N(n+2)+(n-1)]}+\frac{\cal C}{1-{\cal C}}\frac{3n+10}{N(n+2)+(n-1)}~,\\
\label{aslsmall2}
\alpha_{s\ast}&=&-\frac{(11n+20)(n+2)}{4[N(n+2)+(n-1)]^{2}}+\frac{\cal C}{1-{\cal C}}\frac{(3n+10)(n+2)}{[N(n+2)+(n-1)]^{2}}~\nonumber\\
&&-\frac{\cal C}{(1-{\cal C})^{2}}\frac{(3n+10)^{2}}{[N(n+2)+(n-1)]^{2}}~\\
\label{rlsmall2}
r_\ast^{(s)}&=&\frac{8n(1+{\cal C})}{N(n+2)+(n-1)}~
\eea
in the small scale limit. One can see that in both limits, $n_s$, $\alpha_s$ and $r$ are all functions of the power-law index of the potential $n$ and the e-folding number $N_\ast$ only.

\begin{table*}
\centering
\begin{tabular}{c|c|c|c|c|c|c}
\hline
    & \multicolumn{3}{c|}{$N=50$}  &\multicolumn{3}{c}{$N=60$} \\
\cline{2-7}
   &$n=2/3$ &$n=2$ &$n=4$    
   &$n=2/3$ &$n=2$ &$n=4$ \\ 
\hline
    $n_s$   &0.9489  &0.9477   &0.9472  
                 &0.9574  &0.9564   &0.9559 \\ 
\hline
   $\alpha_s$  &-0.0010198  &-0.0010395  &-0.0010456     
                      &  -0.0007088  &-0.0007231  &-0.0007285  \\ 
\hline
   $r$        &0.0399   &0.0796      &0.1056       
                &0.0333   &0.0663      &0.0881   \\  
\hline
\end{tabular}
\caption{The large scale limit $n_s$, $\alpha_s$ and $r$ for large field inflation models with various potential index $n$ and e-folding number $N$.}\label{largelT}
\end{table*}

\begin{figure}
\centering
\includegraphics[scale=0.3]{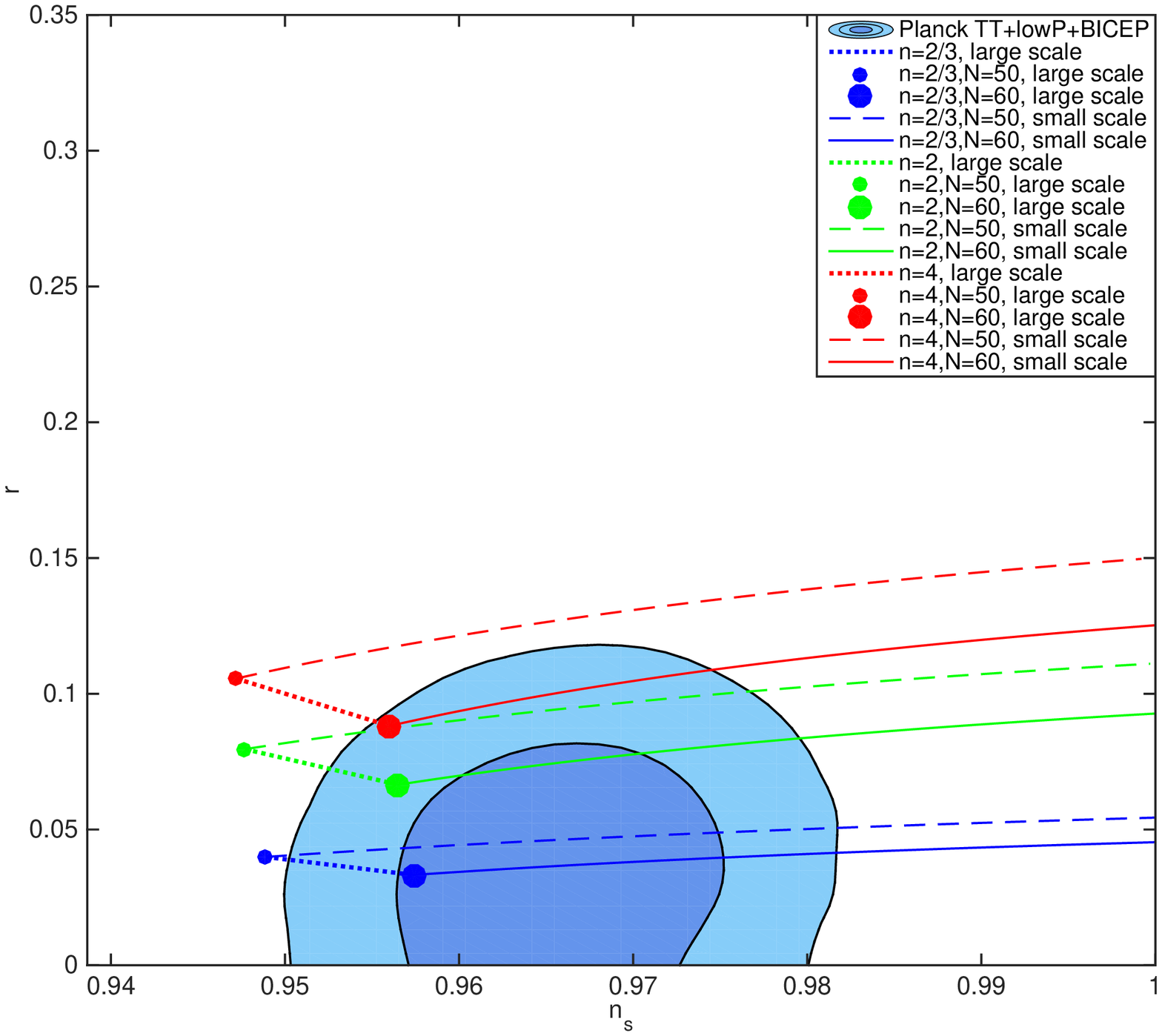}
\includegraphics[scale=0.31]{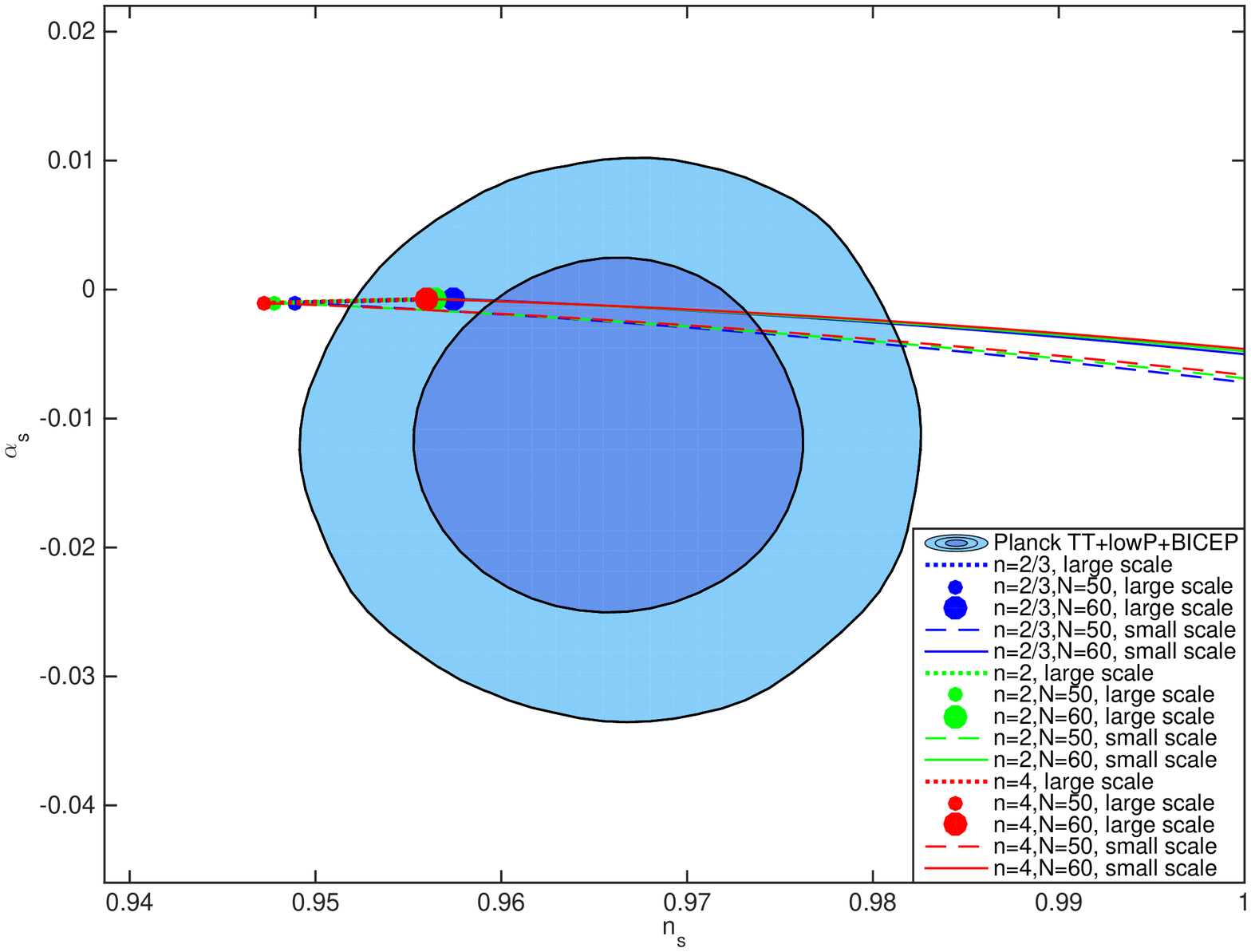}
\caption{Plot of $n_s-r$ (left panel) and $n_s-\alpha_s$ (right panel) of large field inflation models, with comparison to PLANCK TT+lowP+BICEP data.}\label{largeP}
\end{figure}

In Fig. \ref{largeP}, we also plot the constraint contour of $n_s-r$ as well as $\alpha_s-n_s$ for each case, and confront them to the PLANCK TT+lowP+BICEP data \cite{Array:2015xqh}. From the $n_s-r$ plot we can see that most of these cases can fall in the (at least $2\sigma$) confidence level of the observational data. Especially, because of the nonminimally coupling effect, the tensor/scalar ratio of power-law inflation model can be pretty suppressed. The model with smaller $n$ gives lower $r$ and thus has a better fitting to the data, which is similar to GR case, therefore for $n=2/3$ with $N=60$, the model in large scale limit can fit with the data in $1\sigma$ confidence level, while $n>4$ may fall outside the $2\sigma$ region. Moreover, considering small scale can help improve the data fitting. In the $\alpha_s-n_s$ we can see that, both $n_s$ and $\alpha_s$ are not quite sensitive to the parameters, so the data points overlaps with each other. For $N>50$, the data points can fall into $1\sigma$ confidence level.

A side comment on the constraint from Lyth bound is presented below. As D. Lyth suggested in 1996  \cite{Lyth:1996im}, a detectable $r$ will also give rise to a super-Planckian excursion of $\phi$, therefore the effective field theory description of inflation may not be trustable. However, since \cite{Lyth:1996im} only considered canonical large field model and assumed a monotonic slow-roll parameter, many references \cite{Efstathiou:2005tq, BenDayan:2009kv, Choudhury:2013iaa, Antusch:2014cpa, Choudhury:2014wsa, German:2014qza, Gao:2014pca, Hossain:2014ova} argued that for more general inflation models, this bound could be modified. In our model, using Eqs. (\ref{phiel1}), (\ref{phifl1}), (\ref{phiel2}) and (\ref{phifl2}), one can also calculate the excursion of $\phi$ during inflation:
\bea
\Delta\phi&\equiv&|\phi_\ast-\phi_e| \nonumber\\
&=&\left(\frac{M^{2}M_{pl}^{n}}{2\lambda\beta}\right)^{\frac{1}{n+2}}\left([2N(n+2)n+n^{2}]^{\frac{1}{n+2}}-n^{\frac{2}{n+2}}\right)~\text{for~small~$k$}~,\\
& or &\left(\frac{M^{2}M_{pl}^{n}}{\lambda\beta}\right)^{\frac{1}{n+2}}\left([N(n+2)n+(n-1)n]^{\frac{1}{n+2}}-[n(n-1)]^{\frac{1}{n+2}}\right)~\text{for~large~$k$}~.
\eea

From above one can see that, the field variation $\Delta\phi$ depends not only on model parameters such as $n$, $\lambda$, $\beta$, $M$ and the e-folding number $N_\ast$, some of which are not constrained by the observational data. Therefore, it is easy for $\Delta\phi<M_{pl}$ to be satisfied with proper choice of these parameters. For example, if we set the kinetic nonminimal coupling scale $M\sim 10^{-6}M_{pl}$, for $N=50$ inflation with $n=4$, we can get $\Delta\phi\simeq 0.8M_{pl}<1M_{pl}$ for $\lambda=10^{-10}$ and $\beta=1$. A more general analysis can be seen for canonical NKC inflation case in Ref. \cite{Yang:2015pga}.

\subsection{Inflation with Small Field Potential}
In the small field inflation models, the inflaton field goes the opposite direction, namely from a small value towards a large value. Different from large field models, usually for the small field models the initial conditions have to be fine-tuned, and the tensor/scalar ratio is small \cite{Brandenberger:2016uzh}. One possibility of small-field potential can be written as: 
\be\label{smallV}
V(\phi)=\Lambda\bigg[1-\bigg(\frac{\phi}{v}\bigg)^2\bigg]^n~.
\ee
For $|\phi/v|\ll 1$, the potential will reduce to a cosmological constant, $V(\phi)\simeq \Lambda$. For different choices of $n$, several known models can also be recovered. For $n=1$, (\ref{smallV}) is an example of the so-called ``Hill-top" potential \cite{Albrecht:1982wi} and can be used as the potential of the IR model of DBI inflation \cite{Chen:2004gc}, and for $n=2$, (\ref{smallV}) is nothing but the symmetry-breaking potential \cite{Albrecht:1984qt}, which is applied in Higgs-inflation scenario \cite{Bezrukov:2007ep}.  

Consider the small-field potential Eq. (\ref{smallV}) and with Eq. (\ref{srpara}), one can get the potential-based slow-roll parameters as:
\bea
\label{epsilonVs}
\epsilon_{V}&=&\frac{2M^2M_{pl}^4}{\Lambda\beta v^2}\frac{n^2(\phi/v)^2}{[1-(\phi/v)^2]^{n+2}}~,~\\
\label{etaVs}
\eta_{V}&=&-\frac{2M^2M_{pl}^4}{\Lambda\beta v^2}\frac{n[1-(2n-1)(\phi/v)^2]}{[1-(\phi/v)^2]^{n+2}}~,~\\
\label{xiVs}
\xi_V^2&=&-\frac{24M^2M_{pl}^6}{\Lambda^2\beta^2v^4}\frac{{n^2}(n-1)(\phi/v)^2}{[1-(\phi/v)^2]^{2n+3}}+\frac{16M^4M_{pl}^8}{\Lambda^2\beta^2v^4}\frac{{n^2}(n-1)(n-2)(\phi/v)^4}{[1-(\phi/v)^2]^{2n+4}}~.
\eea
Moreover, the e-folding number of inflation evolves from the horizon-crossing to the end, and Eq. (\ref{efold}) will become into:
\bea
\label{Ns}
N&=&\int_{\phi_{*}}^{\phi_{e}}\frac{\Lambda\beta v^2}{2M^2M_{pl}^4}\frac{[1-(\phi/v)^2]^{n+1}}{n\phi}d\phi~\nonumber\\
&=&-\frac{\Lambda\beta v^2}{4M^2M_{pl}^4n}\left(\frac{1-[1-(\phi/v)^2]^{1+n}}{1+n}+(-1)^n B\left[\frac{v^2}{\phi^2},-n,1+n\right]\right)\Bigg|_{\phi_\ast}^{\phi_e}~,
\eea
where $B[v^2/\phi^2,-n,1+n]$ is the incomplete Beta function. 

For $n=1$ case, for common choice of parameters such that $\Lambda=10^{-8}M_{pl}^4$, $\beta=1$, $v=10M_{pl}$ and $M=10^{-6}M_{pl}$, one can find that $\epsilon_V$ reaches unity earlier than $|\eta_V|$, so the final value of $\phi$ has to be chosen at the time when $\epsilon_V=1$. However, according to Eq. (\ref{epsilonVs}), $\epsilon_V(\phi)=1$ is a high order algebraic equation which is difficult to obtain analytical forms of solution, therefore we refer to numerical approach with specific values of parameters. In the above parameter choice, we found that
\be
\label{phies1}
\phi_e\simeq0.993707v=9.93707M_{pl}~.
\ee
Eq. (\ref{Ns}) can be simplified as:
\be\label{Ns1}
N=\frac{\Lambda\beta v^2}{8M^2M_{pl}^4}\left[\frac{\phi^2}{v^2}\left(\frac{\phi^2}{v^2}-4\right)+4\ln\left(\frac{\phi}{v}\right)\right]\Bigg|_{\phi_\ast}^{\phi_e}~,
\ee
therefore, 
\be 
\phi_\ast\simeq0.957786v=9.57786M_{pl}~(N=50)~,~~~\phi_\ast\simeq0.955191v=9.55191M_{pl}~(N=60)
\ee
and one can immediately get $\Delta\phi\simeq0.35921M_{pl}$ ($N=50$) while $\Delta\phi\simeq0.38516M_{pl}$ ($N=60$), both do not violate the Lyth bound. This is due to the fact that $N$ is enhanced by a factor of $M^{-2}$, so in order to get proper $N$, less excursion of $\phi$ is required. By using Eqs. (\ref{epsilonVs}-\ref{xiVs}) as well as the definition of $\epsilon_f$, one gets 
\bea
&&\epsilon_{V\ast}\simeq0.003250~,~\eta_{V\ast}\simeq-0.000293~,~\xi^2_{V\ast}\simeq 0~,~\epsilon_{f\ast}\simeq-0.001171~,~~~(N=50)\\
&&\epsilon_{V\ast}\simeq0.002714~,~\eta_{V\ast}\simeq-0.000261~,~\xi^2_{V\ast}\simeq 0~,~\epsilon_{f\ast}\simeq-0.001042~,~~~(N=60)
\eea
and from Eqs. (\ref{nslsr}), (\ref{aslsr}) and (\ref{rlsr}), we get 
\bea
&&n_{s\ast}^{(l)}\simeq0.948160~,~\alpha_{s\ast}^{(l)}\simeq-0.001031~,~r_\ast^{(l)}\simeq0.052000~,~~~(N=50) \\
&&n_{s\ast}^{(l)}\simeq0.956633~,~\alpha_{s\ast}^{(l)}\simeq-0.000721~,~r_\ast^{(l)}\simeq0.043424~,~~~(N=60) 
\eea
in the large scale limit. For small scale limit,  
from Eqs. (\ref{nsssr}), Eq.(\ref{asssr}) and (\ref{rssr}), we get 
\bea
\label{nsssmall1N50}
n_{s\ast}^{(s)}&\simeq&0.948160+0.139109\frac{\cal C}{1-{\cal C}}~,\\
\label{asssmall1N50}
\alpha_{s\ast}^{(s)}&\simeq&-0.001031+0.002288\frac{\cal C}{1-{\cal C}}-0.013403\frac{\cal C}{(1-{\cal C})^2}~,\\
\label{rssmall1N50}
r_\ast^{(s)}&\simeq&0.052000(1+{\cal C})~
\eea
for $N=50$, which gives the relationship $r^{(s)}=0.104000-0.007234/(n_s^{(s)}-0.809051)$, and 
\bea
\label{nsssmall1N60}
n_{s\ast}^{(s)}&\simeq&0.956633+0.096972\frac{\cal C}{1-{\cal C}}~,\\
\label{asssmall1N60}
\alpha_{s\ast}^{(s)}&\simeq&-0.000721+0.001602\frac{\cal C}{1-{\cal C}}-0.009404\frac{\cal C}{(1-{\cal C})^2}~,\\
\label{rssmall1N60}
r_\ast^{(s)}&\simeq&0.043424(1+{\cal C})~
\eea
for $N=60$, with the relationship $r^{(s)}=0.086848-0.004211/(n_s^{(s)}-0.859661)$, respectively.

One can perform similar calculations for $n=2$. Eq. (\ref{Ns}) can be written as:
\be\label{Ns2}
N=\frac{\Lambda\beta v^2}{4M^2M_{pl}^4}\left[-\frac{3}{2}\left(\frac{\phi}{v}\right)^2+\frac{3}{4}\left(\frac{\phi}{v}\right)^4-\frac{1}{6}\left(\frac{\phi}{v}\right)^6+\ln\left(\frac{\phi}{v}\right)\right]\Bigg|_{\phi_\ast}^{\phi_e}~.
\ee
For parameter choices of $\Lambda$,~$v$ and $M$ as:~$\Lambda=10^{-8}M_{pl}^4$, $\beta=1$, $v=1M_{pl}$, $M=10^{-6}M_{pl}$, we find that $\epsilon_V=1$ can still be used as final condition of inflation, so one gets 
\be
\label{phies2}
\phi_e\simeq0.915991v\simeq0.915991M_{pl}~,
\ee
and
\be
\phi_\ast\simeq0.674107v\simeq0.674107M_{pl}~(N=50)~,~~~\phi_\ast\simeq0.658604v\simeq0.658604M_{pl}~(N=60)
\ee
and $\Delta\phi\simeq0.241884M_{pl}$ ($N=50$) while $\Delta\phi\simeq0.257387M_{pl}$ ($N=60$). The slow-roll parameters are 
\bea
&&\epsilon_{V\ast}\simeq0.002239~,~\eta_{V\ast}\simeq0.001640~,
\nonumber\\
&&~\xi^2_{V\ast}\simeq-0.000030~,~\epsilon_{f\ast}\simeq-0.004926~,~(N=50)\\
&&\epsilon_{V\ast}\simeq0.001925~,~\eta_{V\ast}\simeq-0.001172~,
\nonumber\\
&&~\xi^2_{V\ast}\simeq-0.000119~,~\epsilon_{f\ast}\simeq-0.004407~,~(N=60)
\eea
Therefore from Eqs. (\ref{nslsr}), (\ref{aslsr}) and (\ref{rlsr}), we get
\bea
&&n_{s\ast}^{(l)}\simeq0.957096~,~\alpha_{s\ast}^{(l)}\simeq-0.000503~,~r_\ast^{(l)}\simeq0.035824~,~~~(N=50) \\
&&n_{s\ast}^{(l)}\simeq0.964302~,~\alpha_{s\ast}^{(l)}\simeq0.000135~,~r_\ast^{(l)}\simeq0.030800~,~~~(N=60) 
\eea
in large scale limit, and for small scale limit, from Eqs. (\ref{nsssr}), (\ref{asssr}) and (\ref{rssr}), we get 
\bea
\label{nsssmall2N50}
n_{s\ast}^{(s)}&\simeq&0.957096+0.104012\frac{\cal C}{1-{\cal C}}~,\\
\label{asssmall2N50}
\alpha_{s\ast}^{(s)}&\simeq&-0.000503+0.001104\frac{\cal C}{1-{\cal C}}-0.010818\frac{\cal C}{(1-{\cal C})^2}~,\\
\label{rssmall2N50}
r_\ast^{(s)}&\simeq&0.035824(1+{\cal C})~.
\eea
\begin{figure}
\centering
\includegraphics[scale=0.32]{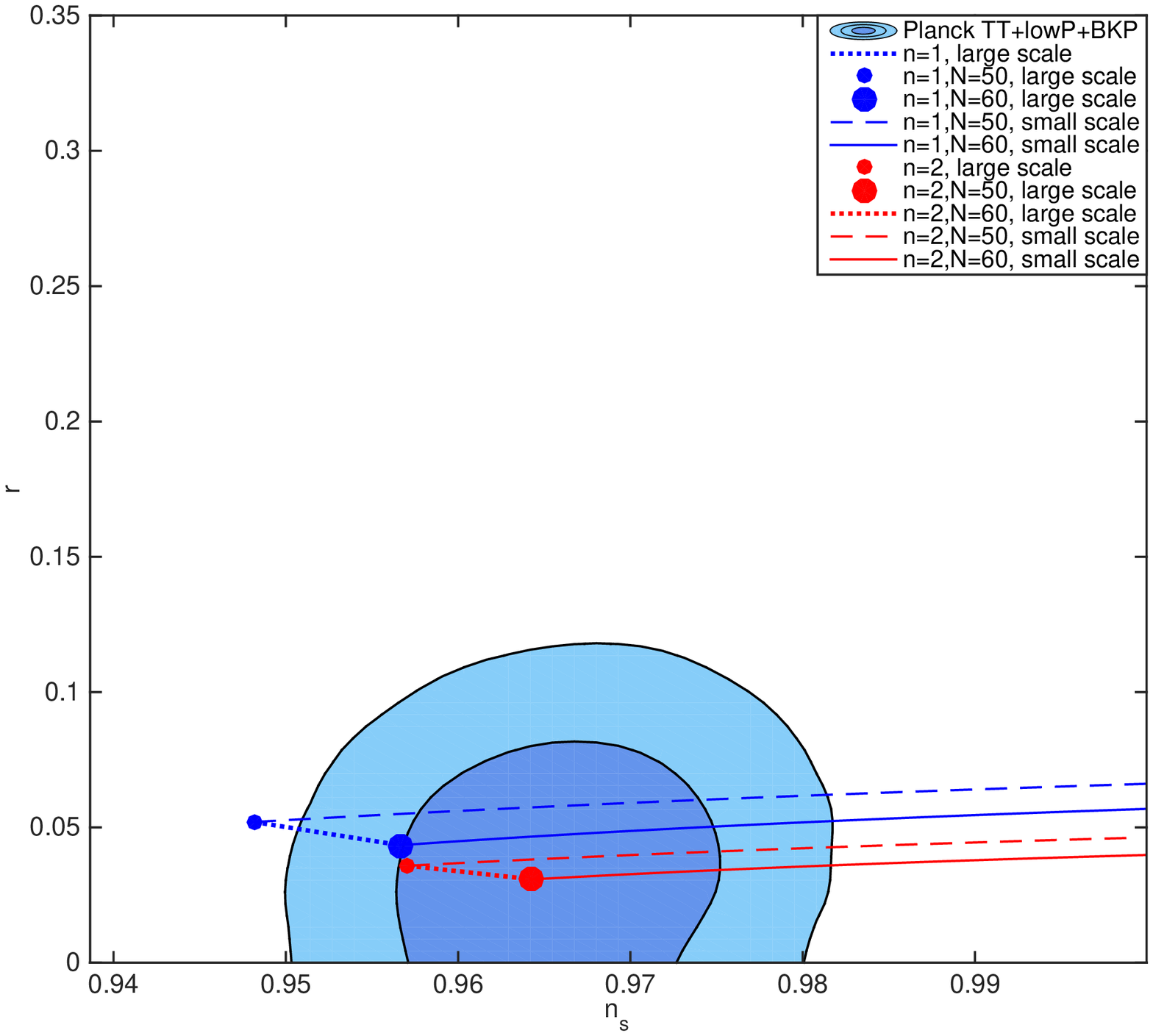}
\includegraphics[scale=0.33]{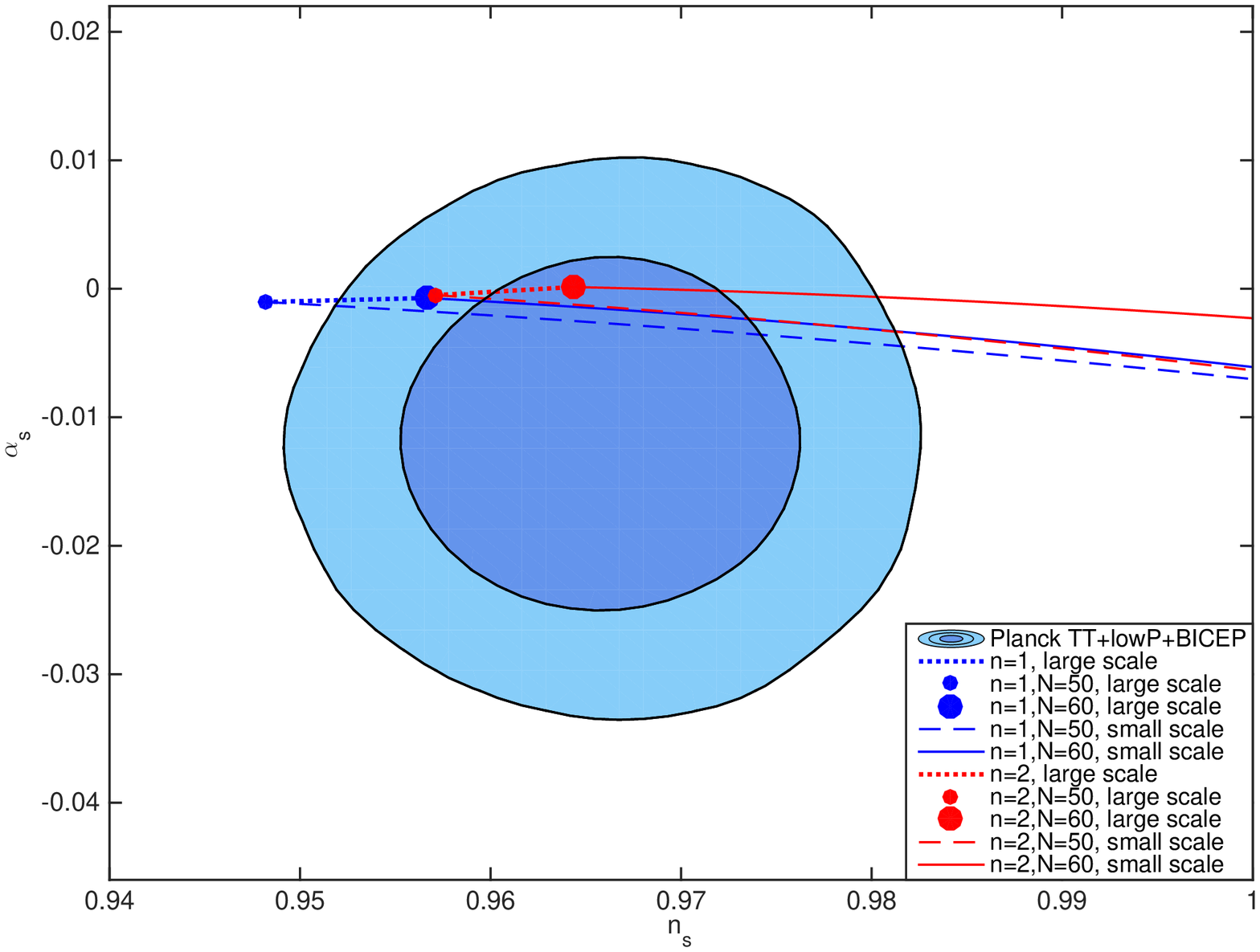}
\caption{Plot of $n_s-r$ (left panel) and $n_s-\alpha_s$ (right panel) of small field inflation models, with comparison to PLANCK TT+lowP+BICEP data.}\label{SmallP}
\end{figure}
for $N=50$, with the relationship $r_\ast^{(s)}\simeq0.071648-0.003726/(n_{s\ast}^{(s)}-0.853084)$, and
\bea
\label{nsssmall2N60}
n_{s\ast}^{(s)}&\simeq&0.964302+0.086265\frac{\cal C}{1-{\cal C}}~,\\
\label{asssmall2N60}
\alpha_{s\ast}^{(s)}&\simeq&0.000135-0.004652\frac{\cal C}{1-{\cal C}}-0.007442\frac{\cal C}{(1-{\cal C})^2}~,\\
\label{rssmall2N60}
r_\ast^{(s)}&\simeq&0.030800(1+{\cal C})~.
\eea
for $N=60$, with the relationship $r_\ast^{(s)}\simeq0.061600-0.002657/(n_{s\ast}^{(s)}-0.878037)$, respectively.

In Fig. \ref{SmallP}, we plot the constraint contour of $n_s-r$ and $\alpha_s-n_s$ for each case, compared to the same data used for large field cases. The results are quite similar. For small field inflation, it is more easy to get smaller $r$ to fit the observational data, and for $n=2$ case, all data points with $N\in [50, 60]$ falls in the $1\sigma$ confidence level in large scale limit. The data of $\alpha_s$ is still not sensitive to the model parameters, and can fit with the data in $1\sigma$ confidence level.

\section{Constraints on Tensor Spectrum: $n_T$ vs. $r$}
\label{ntr}
In this section, we analyse the relationship between spectral index for tensor perturbation $n_T$ and the tensor/scalar ratio $r$ in our model. The future detectability of AliCPT Telescope on $(n_T, r)$ parameter space has been given in \cite{Li:2017drr}. In our model, one has 
\be
s_T\simeq\frac{2}{3}\epsilon_V(3\epsilon_V-\eta_V)~,
\ee
in potential-driven case, which is of the second order of $\epsilon_V$, $\eta_V$, etc. Therefore together with (\ref{relation}) $n_T$, can be expressed as
\be
n_T\simeq -2\epsilon_V+{\cal O}(\epsilon_V^2)~.
\ee
At leading order, $n_T\simeq-2\epsilon_V$. So in large scale limit, the consistency relation of $r=-8n_T$ can be satisfied, while in small scale limit, it is corrected as $r=-8n_T(1+{\cal C})$. However, if we go further to the second or higher orders, $n_T$ will receive a correction from the NKC effect, and we will get a deviation from the consistency relation. Therefore,  if the future observation is sensitive enough to detect second order in $n_T$, one may also tell difference between our model and minimal coupling inflation models.

\begin{figure}
\centering
\includegraphics[scale=0.4]{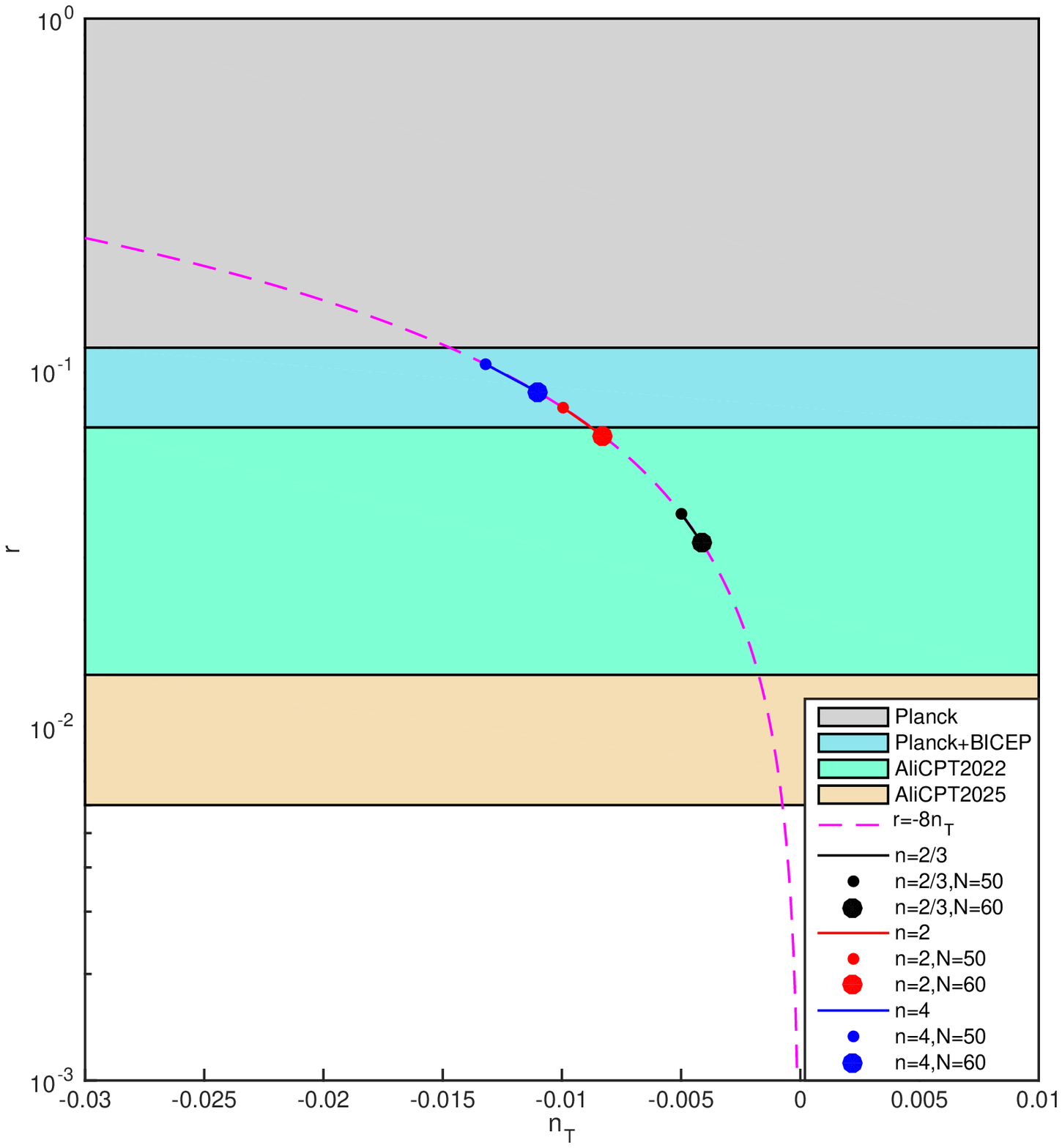}
\includegraphics[scale=0.41]{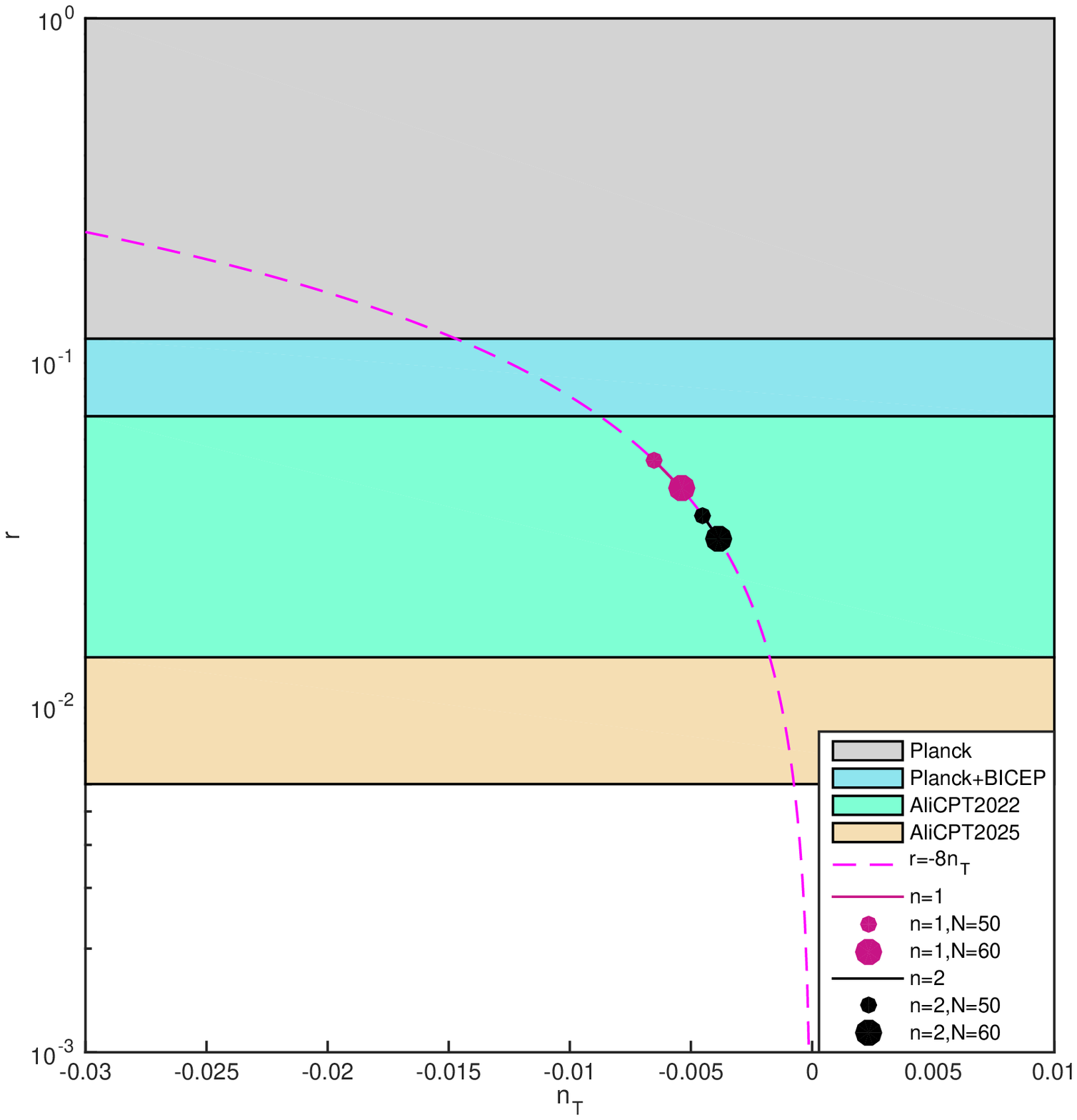}
\caption{$n_T-r$ plots for Large field (left panel) and small field (right panel) respectively.}\label{ntrplot}
\end{figure}

In Figs. \ref{ntrplot}, we plot the relationship of $(n_T, r)$ in large scale limit for both large and small field models, and the constraints from current Planck+BICEP data and future AliCPT predictions. In the figures we can see that, the consistency relation are satisfied in both cases for large scale limit. For large field of $n=2$ and $n=2/3$ and for small field of $n=1$ and $n=2$, the values of $r$ can fall into the detectable region of the next AliCPT predictions, so one can expect to test these cases in near future. For small scale case, the data points will be lifted up due to a factor of $(1+{\cal C})$, with $n_T$ unchanged.

\section{concluding remarks}
\label{conclusion}
In this paper, we have studied the spectral index and its running in Dintic inflation model, mainly focusing on the potential-driven case. In our analysis, we have taken as examples both the large field potential $V(\phi)=\lambda M_{pl}^4(\phi/M_{pl})^n$ and the small field potential $V(\phi)=\Lambda[1-(\phi/v)^2]^n$, with various choices of $n$. We have calculated the power spectrum, spectral index, the running of the index, as well as the tensor spectal index and tensor/scalar ratio. We also have confronted our results to the observational constraints.

From the numerical plots, we can see that in large region parameter choice, the quantities of our model can be acceptable within current observational data. Especially, the tensor/scalar ratio can be suppressed to meet with the data due to the nonminimal coupling term. The constraint on $r$ favors smaller $n$ for large field inflation and larger $n$ for small field inflation, while $\alpha_s$ are not sensitive to model parameters in both cases. Moreover, at small scale region, there is an uncertainty due to the freedom of the parameter ${\cal C}$, so one can give a line rather than a point of $(n_s, r)$ and ($\alpha_s, n_s$), along which one can get some values of parameters which is favored within $1\sigma$ level for the current observational data. We also showed that in leading order, the tensor spectral index still related to tensor/scalar ratio with the famous consistency relation $r=-8n_T$, and we have compared our results with the current and future detecting abilities of the AliCPT telescope for primordial gravitational waves. 

As a final remark, we comment that due to the nonminimal kinetic term and its nonlinear effect, there should be some more differences from GR case, or even the nonminimal kinetic coupling inflation studied in \cite{Yang:2015pga}, For example, at subleading order, the consistency relationship of $(n_T, r)$ can get modified. However, these differences are still very small (several orders of slow-roll parameters), and undetectable within the current observational data. We hope the future observations with higher precisions and detectability can give us a better distinction of different models, and a better test of our model.

\begin{acknowledgments}
We thank Ze Luan and Shulei Ni for help with plots drawing. This work is supported in part by NSFC under Grant No: 11405069, 11653002, 11735007 and 111375070.

\end{acknowledgments}

\end{document}